\documentstyle[twocolumn,aps]{revtex}
\begin{document}
\title{
Comment on ``Suppression of $\pi NN^*$ Coupling\\ and Chiral Symmetry"}
\author{Michael C. Birse}
\address{Theoretical Physics Group, Department of Physics and Astronomy,\\
University of Manchester, Manchester, M13 9PL, U.K.\\}
\maketitle

In a recent Letter\cite{joh98}, Jido, Oka and Hosaka have suggested that the
$\pi NN^*$ coupling of the odd-parity $N(1535)$ resonance is strongly
suppressed as a consequence of chiral symmetry. Their argument is based on a
pion-to-vacuum correlator of two interpolating fields,
\begin{equation}\label{correl}
\Pi^a(p) = i \int\! d^4x\, e^{ip\cdot x} \langle 0| T\bigl(J(x;s), \overline 
J(0;t)\bigr) |\pi^a(q)\rangle,
\end{equation}
where the baryon fields $J(x,s)$ are taken from a one-parameter family of
local products of three quark fields, with $J(x,-1)$ being the Ioffe 
field\cite{iof81}. Jido, Oka and Hosaka consider the $q\rightarrow 0$ limit
of this correlator, and use the soft-pion theorem to express it as a chiral
rotation of the vacuum-to-vacuum correlator of the same baryon fields. The
chiral transformation properties of their fields then imply that
the correlator (\ref{correl}) is purely proportional to $\gamma_5$, with no
piece of the form $p\llap/\gamma_5$.

The contribution of the $N(1535)$ to the spectral representation
of (\ref{correl}) for $q\rightarrow 0$ contains a term of the form
\begin{eqnarray}\label{piNNstar}
g_{\pi NN^*}&&[\lambda_N(t)\lambda_{N^*}(s)-\lambda_N(s)\lambda_{N^*}(t)]
\nonumber\\
&&\times{m_N+m_{N^*}\over (p^2-m_N^2)(p^2-m_{N^*}^2)}ip\llap/\gamma_5,
\end{eqnarray}
in the narrow resonance approximation. Here the $\lambda$'s denote the
coupling strengths of the interpolating fields to the baryon states, and
$g_{\pi NN^*}$ is the coupling of interest. The fields are chosen with $s\neq
t$, such that one couples strongly to the nucleon, and the other to the
$N(1535)$. If this contribution were the only one with this Dirac structure,
then the coupling $g_{\pi NN^*}$ would vanish, as claimed in\cite{joh98}. 

Note that this result for $g_{\pi NN^*}$ relies crucially on the soft pion
limit, $q\rightarrow 0$, and hence at least one of the baryon states must be
off-shell. This means that for any $p^2$ there will be contributions from all
states that can be created by the corresponding interpolating fields. If the
spectral representation of the correlator (\ref{correl}) is approximated by a
set of discrete states, then the $p\llap/\gamma_5$ piece can be written as a
sum of terms of the form (\ref{piNNstar}), where the sum runs over all pairs
of spin-${1\over 2}$ baryon states. By rewriting each term as a difference of
single poles, this sum can be expressed in the form
\begin{equation}\label{ressum}
\sum_\alpha{ip\llap/\gamma_5\over p^2-m_\alpha^2}
\sum_{\beta\neq\alpha}A_{\alpha\beta},
\end{equation}
Here I have defined 
\begin{equation}
A_{\alpha\beta}=g_{\pi\alpha\beta}[\lambda_\alpha(t)
\lambda_\beta(s)-\lambda_\alpha(s)\lambda_\beta(t)]/|m_\alpha\pm m_\beta|,
\end{equation}
where the sign in the denominator is determined by the relative parity of the
two states.

Chiral symmetry requires that this sum vanish for all values of $p^2$,
imposing the constraints $\sum_\beta A_{\alpha\beta}=0$ on the couplings. The
antisymmetry of $A_{\alpha\beta}$ means that, for a spectrum with $n$ states,
we have $n-1$ constraints on $n(n-1)/2$ quantities. If only two states, $N$ and
$N^*$, contribute then this does require $A_{NN^*}=0$, from which the result
of\cite{joh98} follows: either $g_{\pi NN^*}$ vanishes or at least one of the
states does not couple to the chosen fields. In contrast, if more than two
states contribute, the vanishing of (\ref{ressum}) does not demand that the
individual $A_{\alpha\beta}$'s vanish.

Nonetheless one might hope that a good choice of the parameters $s$ and $t$
might minimise the couplings of the fields to other states. If such couplings
were indeed small, one could then conclude that $g_{\pi NN^*}$ is suppressed.
However, far from being small, the contributions of baryon continuum states to
correlators of this type are in fact divergent. As stressed by
Ioffe\cite{is84}, the spectral representation of the correlator (\ref{correl})
should be expressed in the form of a double dispersion relation, with
subtractions to remove the UV divergences. These subtractions generate terms
with poles at either $p^2=m_N^2$ or $p^2=m_{N^*}^2$, and which have unknown
coefficients. It is impossible to disentangle such terms from the one of
interest (\ref{piNNstar}) by going to one of the baryon poles. Moreover the
$p$-dependence of these terms means that they cannot be eliminated by the
usual Borel transform used in QCD sum rules.

Hence the absence of a $p\llap/\gamma_5$ piece from correlator (\ref{correl})
is a statement about the particular combination of pion-baryon couplings and
subtraction terms that corresponds to the chosen interpolating fields. It
does not imply that the physical $\pi NN^*$ coupling of the $N(1535)$ is
suppressed as a consequence of chiral symmetry alone. For example, sum rules
based on a different choice of field for the $N(1535)$ show no evidence for
suppression of $g_{\pi NN^*}$ compared to $g_{\eta NN^*}$\cite{kl97}.

This work was supported by the EPSRC.

\end{document}